\documentclass{article}

\usepackage{PRIMEarxiv}

\usepackage[utf8]{inputenc} 
\usepackage[T1]{fontenc}    
\usepackage{hyperref}       
\usepackage{url}            
\usepackage{booktabs}       
\usepackage{amsfonts}       
\usepackage{nicefrac}       
\usepackage{microtype}      
\usepackage{lipsum}
\usepackage{fancyhdr}       
\usepackage{graphicx}       
\graphicspath{{media/}}     

\usepackage{float}
\usepackage{subfig}
\usepackage{dblfloatfix}
\usepackage{color}
\usepackage{tabularx}
\usepackage{adjustbox}
\usepackage{graphicx}
\usepackage{xfrac}
\usepackage{comment}
\usepackage{ulem}

\title{Fast Autofocusing using Tiny Transformer Networks for Digital Holographic Microscopy
}

\author{
  Stéphane Cuenat, Louis Andréoli, Antoine N.André, Patrick Sandoz,\\ \textbf{ Guillaume J. Laurent, Raphaël Couturier and Maxime Jacquot},\\
  Institut FEMTO-ST, \\
 CNRS \& Université Bourgogne Franche-Comté \\
  France \\
  \texttt{stephane.cuenat@univ-fcomte.fr} \\
}

\begin{document}
\maketitle

\begin{abstract}
The numerical wavefront backpropagation principle of digital holography confers unique extended focus capabilities, without mechanical displacements along z-axis. 
However, the determination of the correct focusing distance is a non-trivial and time consuming issue. 
A deep learning (DL) solution is proposed to cast the autofocusing as a regression problem and tested over both experimental and simulated holograms.
Single wavelength digital holograms were recorded by a Digital Holographic Microscope (DHM) with a 10$\mathrm{x}$ microscope objective from a patterned target moving in 3D over an axial range of 92~µm.
Tiny DL models are proposed and compared such as a tiny Vision 
Transformer (TViT), tiny VGG16 (TVGG) and a tiny Swin-Transfomer (TSwinT). The proposed tiny networks are compared with their original versions (ViT/B16, VGG16 and Swin-Transformer Tiny) and the main neural networks used in digital holography such as LeNet and AlexNet.
The experiments show that the predicted focusing distance $Z_R^{\mathrm{Pred}}$ is accurately inferred with an accuracy of 1.2~µm in average in comparison with the DHM depth of field of 15~{\textmu}m. 
Numerical simulations show that all tiny models give the $Z_R^{\mathrm{Pred}}$ with an error below 0.3~{\textmu}m. 
Such a prospect would significantly improve the current capabilities of computer vision position sensing in applications such as 3D microscopy  for life sciences or micro-robotics. 
Moreover, all models reach an inference time on CPU, inferior to 25~ms per inference.
In terms of occlusions, TViT based on its Transformer architecture is the most robust.
\end{abstract}


\section{Introduction}

One major drawback when 3D moving samples are studied in microscopy is the balance between the focal range that limits out-of-plane measurements and the requirement of a high axial resolution, i.e. a short depth of field (DoF)  (see for example \cite{sun2004autofocusing, geusebroek2000robust}).
Several solutions have been proposed such as depth-from-focus imaging \cite{xiong1993depth} and confocal microscopy \cite {kino1996confocal} to reconstruct a topography of the scene. 
Scanning electron microscopy \cite{leamy1982charge} can also be used to get  large in-focus depths. 
In any case, all these methods require a scanning of the scene that slows down the image acquisition rate. 
Moreover, the working distances of these devices are very short and this reduces considerably the interest of a contactless measurement.
\\
\indent Coherent imaging approaches such as Digital Holography (DH) can be used instead of conventional microscopy to address the focusing issues~\cite{Dubois:06,Ferraro:05}.
DH offers a means for recording the phase and amplitude of a propagating wavefront on a solid-state image sensor \cite{Schnars:94}.
Then, by numerically propagating the recorded wavefront backward or forward at particular distances of interest, different characteristics can be extracted, typically three-dimensional surfaces, but also polarization states and intensity distributions. 
Several recording and processing schemes have been developed to assess diverse optical characteristics that make DH a highly powerful coherent imaging method for metrological applications \cite{cuche1999digital, jacquot2001high}.
Significant progress, potential impact and challenging issues in the field of DH can be found in this recent roadmap article \cite{javidi2021roadmap}.
\\
\\
The targeted application aims to address the 3D position measurement needs encountered in small-scale mechatronics \cite{zhang2019robotic, andre2020sensing}. At the microscale, automation involves centimeter-sized actuators necessary to perform diverse tasks with a high accuracy, down to the nanometer range. 
Contactless sensors are thus desired to control 3D motions with a high accuracy over large ranges~\cite{zhang2019robotic}. 
Combined with optical microscopy, computer vision constitutes an efficient means to detect in-plane position and displacements. 
However, microscope objective (MO) lenses provide short in-focus depths and inherently rely on mechanical displacements along the optical axis. 
One way to extend computer vision capabilities to 3D microscopy is to harness the wave character of light by means of DH. 
DH is particularly suited to this aim because it requires a single hologram to reconstruct a 3D scene and because digital back-propagation computations allows image reconstruction in a extended in-focus depths  \cite{colomb2010extended}.
A key-point in DH is to note that, instead of an image of the object, it is the propagating wavefront incident on the image sensor that is recorded. 
The distance of the object does not impact the recording quality, it only changes the actually recorded wavefront in accordance with scalar diffraction theory. 
Therefore, blur does not exist at the recording stage of DH since it does not seek for any in-focus image. 
The object distance stands for a computation parameter that is numerically tunable over an extended range, but limited to coherent length of the light source used. 
This specificity makes the range of working distances allowed by DH incomparably larger than that allowed by conventional incoherent imaging methods.
DH can be applied to micro-objects in microscopy with a Digital Holographic Microscope (DHM) setup. 
The reconstruction distance leading to the best-focused image has to be determined among the z-range explored by the object. 
There are various techniques for defining image-formation sharpness-criteria that apply to DH \cite{fonseca2016comparative, langehanenberg2008autofocusing}.
\\
\indent Recently, many studies have proposed to study the capabilities of deep-learning Convolutional Neural Network (CNN) to determine in DH various unknown parameters such as focusing distance, or the phase recovery \cite{ren2018learning,rivenson2018phase, wu2018extended, zhang2018fast, pitkaaho2019focus, Zeng:21}. 
These works have to be considered in the wider context of imaging techniques where Deep Learning (DL) approaches are applied to solve complex problems found in computer vision as well as in microscopy \cite{lecun2015deep,sinha2017lensless,jiang2018transform}.
A recent work~\cite{ren2018learning} even demonstrated that Deep CNN gives better results in terms of prediction of propagation distance in DH without knowing all the setup's physical parameters, than other learning-based algorithms such as Multi Layer Perceptron (MLP) \cite{haykin1994neural}, support vector machine \cite{cortes1995support}, and k-nearest neighbor \cite{Mucherino2009}. The
hardware implementation of artificial neural networks has constituted a real challenge for many years \cite{psaltis1995holography, larger2017high, lin2018all}, but the tasks that can be solved by such systems are limited to standard tests of classification and prediction, and they are still limited in terms of scalability for mega-pixel image processing.

This paper aims to illustrate a new high-profile application of machine learning by elevating DHM and autofocusing to a new level. 
Whereas many studies focus on life science microscopy \cite{rivenson2018phase,Pinkard:19, Zeng:21}, this work explores extended visual capabilities offered by combining DH and last generation of DL algorithms such as Vision Transformer (ViT)\cite{dosovitskiy2021image} and Swin-Transformer (SwinT)\cite{liu2021Swin} networks for applications
in micro-robotics \cite{zhang2019robotic, andre2020sensing} or in real-time 3D microscopy \cite{Pinkard:19}. This work introduces for the first time the neural network Transformer architectures applied to advanced coherent imaging field, such as digital holography. This is significant because these new generations of algorithms have already revolutionized the Natural Language Processing (NLP) and recent versions ViT \cite{dosovitskiy2021image} and SwinT\cite{liu2022convnet} highly perform for image recognition thanks to their self-attention feature\cite{vaswani2017attention}. More specifically, our work deals with these new generation of deep learning approaches for autofocusing in digital holographic microscopy to obtain in-focus depth prediction with high accuracy. We developed new tiny ViT and tiny SwinT network architectures, and compared them with typical Convolutional Neural Network (CNN) ones used in optics and digital holography such as AlexNet\cite{pitkaaho2019focus}, VGG\cite{simonyan2015deep} and LeNet\cite{Jaferzadeh:19}.
Swin-Transformers propose a hierarchical Transformer whose representation is computed with Shifted windows. 
The shifted windowing scheme brings greater efficiency by limiting self-attention computation to non-overlapping local windows while also allowing cross-window connection.
This hierarchical architecture has the flexibility to model at various scales and has linear computational complexity with respect to the size of images. 
Taking into account the demand for real time application and achievable training with a reasonable amount of data, tiny networks are developed. 
These first results pave the way to overcome in-focus depth limit\cite{rivenson2018phase} with a short DoF in DHM without any MO lens mechanical displacements\cite{Pinkard:19}.

\section{New trends in deep learning for image processing: ViT and SwinT}

\begin{figure*}[h!]
    \centering
    \includegraphics[width=0.9\linewidth]{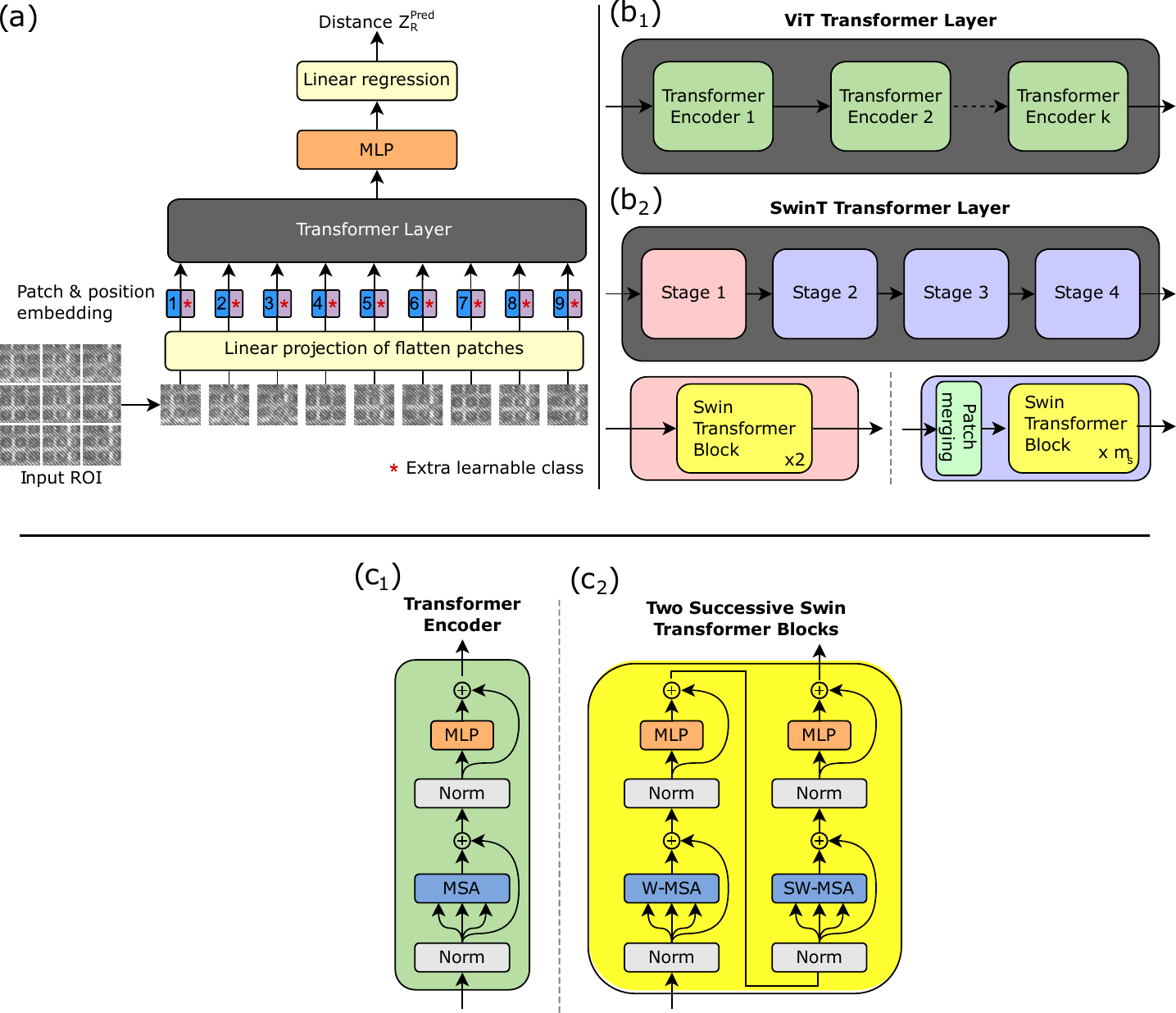}
    \caption{
    Schematic illustration of the ViT \cite{dosovitskiy2021image} and SwinT \cite{liu2021Swin} architectures.
    (a) gives the general architecture of a ViT or a SwinT. 
    The region of interest (ROI) is divided into patches which are linearly projected on an embedded dimension followed by a transformer layer, MLP and the linear regression. 
    (b1) shows the successive layers for a ViT respectively k Transformer Encoders. 
    (b2) gives a view of the SwinT Transfomer layer architecture which is formed by a series of 4 Stages. 
    Each Stage encapsulates a Swin-Transformer Block which is repeated $m_s$ times (with $s$ representing the stage number) and patch merging layers (for the stages 2 to 4). 
    Through the stages, the multi-head attention is computed taking different sizes of patch size and shifted windows. 
    (c1) and (c2) represent the internal steps inside a Transformer Encoder and the two successive Swin-Transformer Block, where the Multi-Self Attention (MSA) is computed (MSA, W-MSA and SW-MSA). 
    SwinT first computes a window multi-head attention (W-MSA) and then a shifted windows multi-head attention (SW-MSA). 
    } 
    \hrulefill
    \label{fig:tiny_vit_architecture}
\end{figure*}

Since the inception of DL neural networks, CNN occupies the field with architectures like VGG-16 \cite{simonyan2015deep}, Densenet \cite{huang2018densely} or EfficientNet \cite{tan2020efficientnet}. 
At the core of a CNN, there is a series of mixed convolution and pooling layers which extract a set of features from an input image. 
One of the main advantages of a CNN compared to MLP Network (first network proposed), is that they are translation invariant and less demanding in resource when it comes to large inputs. 
Later the ViT architecture was introduced. 
This architecture is based on the concept of attention \cite{vaswani2017attention}. 
The attention mechanism was born to help memorize long source sentences in NLP. 
Rather than building a single context vector out of the encoder’s last hidden state, the attention creates shortcuts between the context vector and the entire source input. 
The weights of these shortcut connections are customizable for each output element. 
\\
\\
ViT brings this concept to computer vision \cite{dosovitskiy2021image}. 
ViT is a pure Transformer architecture built from Transformer encoder layers to approach a classification or regression problem. 
ViT splits an input image in a series of patches which would be treated as word tokens by a Transformer Network. 
SwinT even surpasses the performance of a pure ViT network as shown in \cite{liu2022convnet}. 
SwinT extends a ViT network by varying the patch dimension and computes the attention only for a given window (shifted-window over the space of the input image). 
Such a network is able to better assess the local and global information inside the input image.

Figure~\ref{fig:tiny_vit_architecture} shows the architecture of a ViT \cite{dosovitskiy2021image} and SwinT \cite{liu2021Swin} network. 
Panel~in Fig.~\ref{fig:tiny_vit_architecture}(a) gives the global architecture of a ViT or SwinT network and in particular how the Region Of Interest (ROI) is processed. 
This input image is split into different patches and projected on an embedded dimension through the linear projection of the flatten patches (tokens). 
An additional position embedding and class token are added. 
The class token is the only token used to apply a regression. 
Each embedded patch is processed by the transformer layer which  outputs the associated class token after the MLP block. 
The regression layer projects the class token to a scalar, the reconstruction distance~Z in our case.
Panels~in Fig.~\ref{fig:tiny_vit_architecture} ($b_1$) and ($b_2$) describe in more details the Transformer Layer (in gray)  for ViT and SwinT models, respectively.
For ViT, there is a total of k Transfomer Encoder layers.
In the SwinT architecture, the Transformer Layer is composed of a series of $s$ stage layers with typically $s=4$.
The first stage layer contains a two Swin-Transformer block.
The $s-1$ other stages encapsulate a patch merging and $m_s$ Swin-Transformer Blocks, where $m_s$ can change through the stages $s$ (typically $m_s \in [1,4]$). 
The window size is set to a fix value (default: $7\mathrm{x}7$ patches).  
Moreover, the patch size of each stage is increased by a factor 2 through the patch merging layers \cite{liu2021Swin}.  
This creates a hierarchy in comparison to a ViT which always considers the same patch size and a global window \cite{dosovitskiy2021image}. 
The Transformer Encoder (in green) of the ViT and the Swin Transformer block (in yellow) are explained in more details in panels of Fig.~\ref{fig:tiny_vit_architecture}($c_1$) and ($c_2$), respectively.
The input of the transformer encoder is first normalized.
The Multi-head Self-Attention (MSA) is first computed and then followed by a normalization and a MLP block.
In the case of a SwinT, the architecture is similar but with two successive Swin-Transformer blocks where the MSA is first a Window Multi-head Self-Attention (W-MSA), then a Shifted Window Multi-head Self-Attention (SW-MSA). 
While the MSA (ViT case) is computed on the complete set of patches, the W-MSA (SW-MSA) uses a dedicated (shifted) window (SwinT case). SW-MSA allows inter-window interactions.

\subsection{Multi-head self-attention}
\begin{figure}[t!]
    \centering
    \includegraphics[width=200pt]{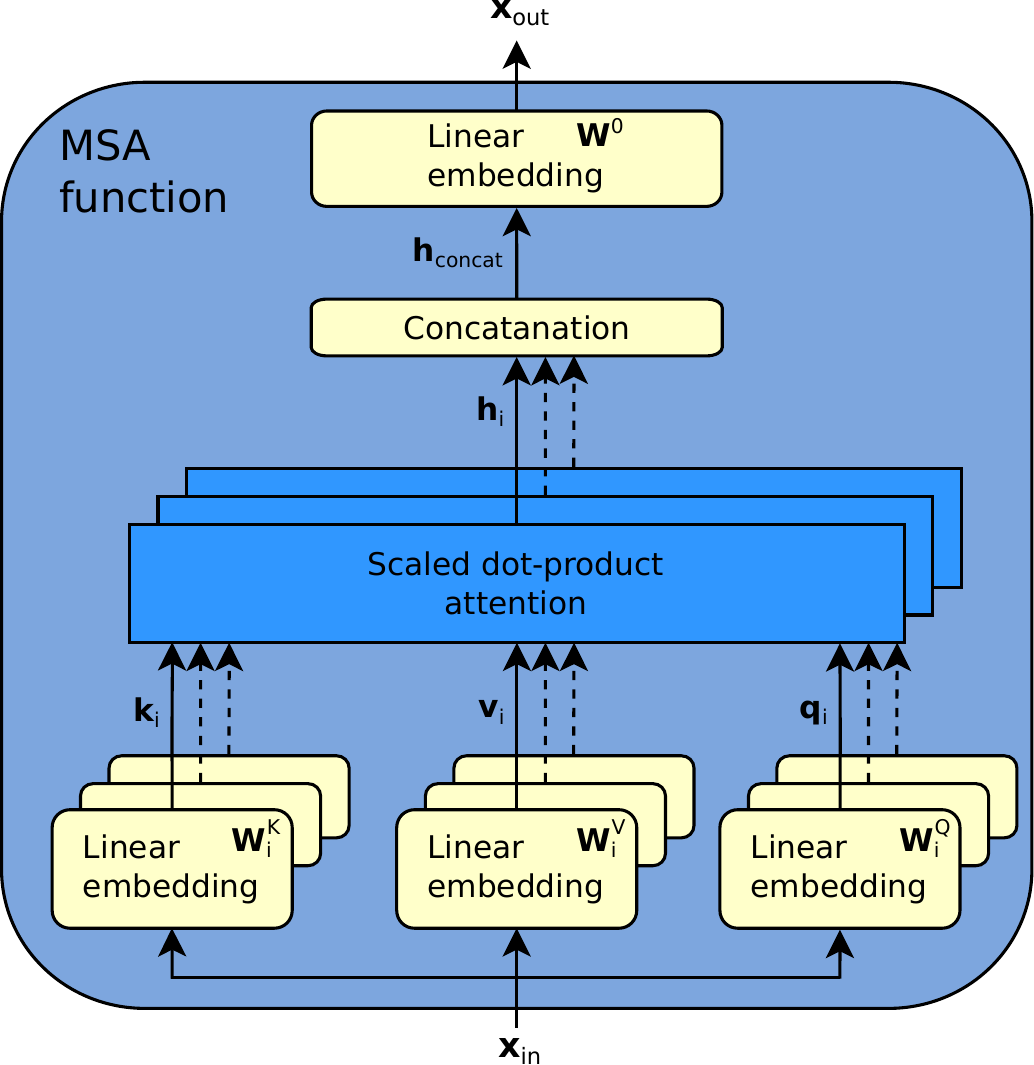}
    \caption{Multi-Headed Self-Attention implemented as a scaled dot-product attention. MSA, W-MSA and SW-MSA blocks on Fig.~\ref{fig:tiny_vit_architecture}.
    }
    \hrulefill
    \label{fig:multi_head_attention}
\end{figure}
As schematically illustrated in Fig.~\ref{fig:multi_head_attention}, the MSA function is approached from a general perspective where the vector $x_\mathrm{in}$ is its input and $x_\mathrm{out}$ its output. 
For each $i\in[1,N] $, the head $\mathbf{h}_i$ is implemented as a scaled dot-product attention.
The N heads $\mathbf{h}$ is called Multi-head Self-Attention.
In case of ViT, $N$ is fixed for all the Transformer Encoder layers. 
SwinT  defines a $N$ for each Stage.
A key $\mathbf{k}_i$, value $\mathbf{v}_i$ and query $\mathbf{q}_i$ dimensional vectors are computed for each head $\mathbf{h}_i$  by projecting the input $\mathbf{x}_\mathrm{in}$ using three learnable matrices ($\mathbf{W}^K_i$, $\mathbf{W}^V_i$, $\mathbf{W}^Q_i$): 
\begin{equation}  \label{eq:5}
    \mathbf{k}_i = \mathbf{W}^K_i
    \mathbf{x}_{\mathrm{in}},
\end{equation}
\begin{equation}  \label{eq:6}
    \mathbf{v}_i =  \mathbf{W}^V_i
    \mathbf{x}_{\mathrm{in}},
\end{equation}
\begin{equation}  \label{eq:7}
    \mathbf{q}_i = \mathbf{W}^Q_i
    \mathbf{x}_{\mathrm{in}}.
\end{equation}
For each head $\mathbf{h}_i$, the attention is computed by taking the key, value and query vectors.
\begin{equation}  \label{eq:4}
    \mathbf{h}_i = \text{Attention}(\mathbf{k}_i, \mathbf{v}_i, \mathbf{q}_i).
\end{equation}
The attention is calculated by first applying a Softmax \cite{miranda2017softmax} used to normalize the dot product between a vector of keys $\mathbf{k}_i$ and a vector of queries $\mathbf{q}_i$. Subsequently, this output acts as weights for the value vector $\mathbf{v}_i$, hence
\begin{equation} \label{eq:8}
    \text{Attention}(\mathbf{k}_i, \mathbf{v}_i, \mathbf{q}_i) = \text{Softmax}(\frac{\mathbf{q}_i\mathbf{k}_i^T}{\sqrt{D}})\mathbf{v}_i,
\end{equation}
where $D$ is the dimension of the key and query vectors ($\mathbf{k}_i$ and $\mathbf{q}_i$).
All the heads $\mathbf{h}_i$ are concatenated as
\begin{equation}  \label{eq:2}
    \mathbf{h}_{\mathrm{concat}} = \mathrm{Concat}(\mathbf{h}_1, \mathbf{h}_2, ..., \mathbf{h}_N).
\end{equation}
The output vector $x_\mathrm{out}$ is obtained by the vector product of $\mathbf{h}_{\mathrm{concat}}$ and a learnable matrix $\mathbf{W}^0$ as 
\begin{equation}  \label{eq:3}
    \mathbf{x}_\mathrm{out} = \mathbf{W}^0 \, \mathbf{h}_{\mathrm{concat}}.
\end{equation}
Multi-head attention is used since it allows the network to attend to different learned representations at different regions of input ROI as described by Fig.~\ref{fig:multi_head_attention} and expressed as 
\begin{equation}  \label{eq:1}
    \mathbf{x}_\mathrm{out} = \mathrm{MSA}(\mathbf{x}_\mathrm{in}). 
\end{equation}

\subsection{Tiny networks: TViT, TSwinT \& TVGG}
In this paper, tiny versions of the original ViT (TViT), Swin-Transformer (TSwinT) and VGG16 (TVGG) are proposed. Tiny networks allow to reduce the number of parameters without impacting much the accuracy of the models. Moreover, tiny models need less computation power and approach real-time processing. Figure~\ref{fig:tiny_vit_architecture} gives an overview of a ViT and SwinT architectures.
TViT modifies a ViT as follows: is built with a total of 12 Transformer Encoder, 8 heads and a patch size of 16x16. The hidden-size of the Transformer encoder has been reduced from 768 (for a ViT/B16) to 128 (TViT) hidden neurons. The MLP dimension (Transformer Encoder) has been reduced to 1024 instead of 3072 hidden neurons (for a ViT/B16).
TSwinT is a revisit of the SwinT architecture where several changes have been applied: the size of the embedding vector is set to 32, the number of Swin-Transformer blocks has been set as follows for each Stage: $m_1 = 2$, $m_2 = 2$, $m_3 = 4$, $m_4 = 2$. The number of heads for each Stage has also been modified: $N_1 = 2, N_2 = 4, N_3 = 8, N_4 = 8$. The window size has been fixed to 4 with an initial patch size of 4x4. 
TViT and TSwinT contrasts with canonical ViT architectures as these models are usually able to learn high-quality intermediate representations with large amounts of data as described in \cite{Zeng:21,raghu2021vision}. 
TVGG is introduced to reduce the number of parameters of the original VGG16 \cite{simonyan2015deep} architecture for comparison purposes.
All filters of each 2D convolution layer have been divided by 2 inside a TVGG architecture.
These changes limit the width of the layers of the tiny networks by keeping their capacity to learn. The number of parameters for each model has drastically diminished as shown in table~\ref{tab:parameters}, by a factor between 5 and 20.
Moreover, all models are trained from scratch only using experimental or simulated digital holograms of different patterns (pseudo-periodic pattern and USAF pattern) without any transfer learning from a pre-trained model on a dataset like ImageNet \cite{deng2009imagenet}.  
The tiny models take as input ROI of 128x128 of a single wavelength digital hologram.

\begin{table}[h!]
    \centering
    \begin{adjustbox}{max width=\columnwidth}
    \begin{tabular}{c|c|c|c|c|}
        \multicolumn{2}{c|}{\textbf{Tiny model}}  & \multicolumn{3}{c|}{\textbf{Original model (pre-trained)}}\\
        \cline{1-5}
     \textbf{Model} & \textbf{\# parameters} & \textbf{Model} & \textbf{\# parameters} & \textbf{Ref}\\
     \hline
        TVGG & $3 \cdot 10^{6}$ & VGG16 & $14 \cdot 10^{6}$ & \cite{simonyan2015deep}\\
        TViT &  $4 \cdot 10^{6}$ & ViT-B16 & $85 \cdot 10^{6}$ & \cite{dosovitskiy2021image}\\ 
        TSwinT &  $2.7 \cdot 10^{6}$ & SwinT Tiny & $28 \cdot 10^{6}$ & \cite{liu2021Swin} \\
    \end{tabular}
    \end{adjustbox}
    \caption{Number of parameters for each tiny neural network compared to the original version.
    }
    \hrulefill
    \label{tab:parameters}
\end{table}

\begin{figure}[b!]
    \centering
    \includegraphics[width=0.88\linewidth]{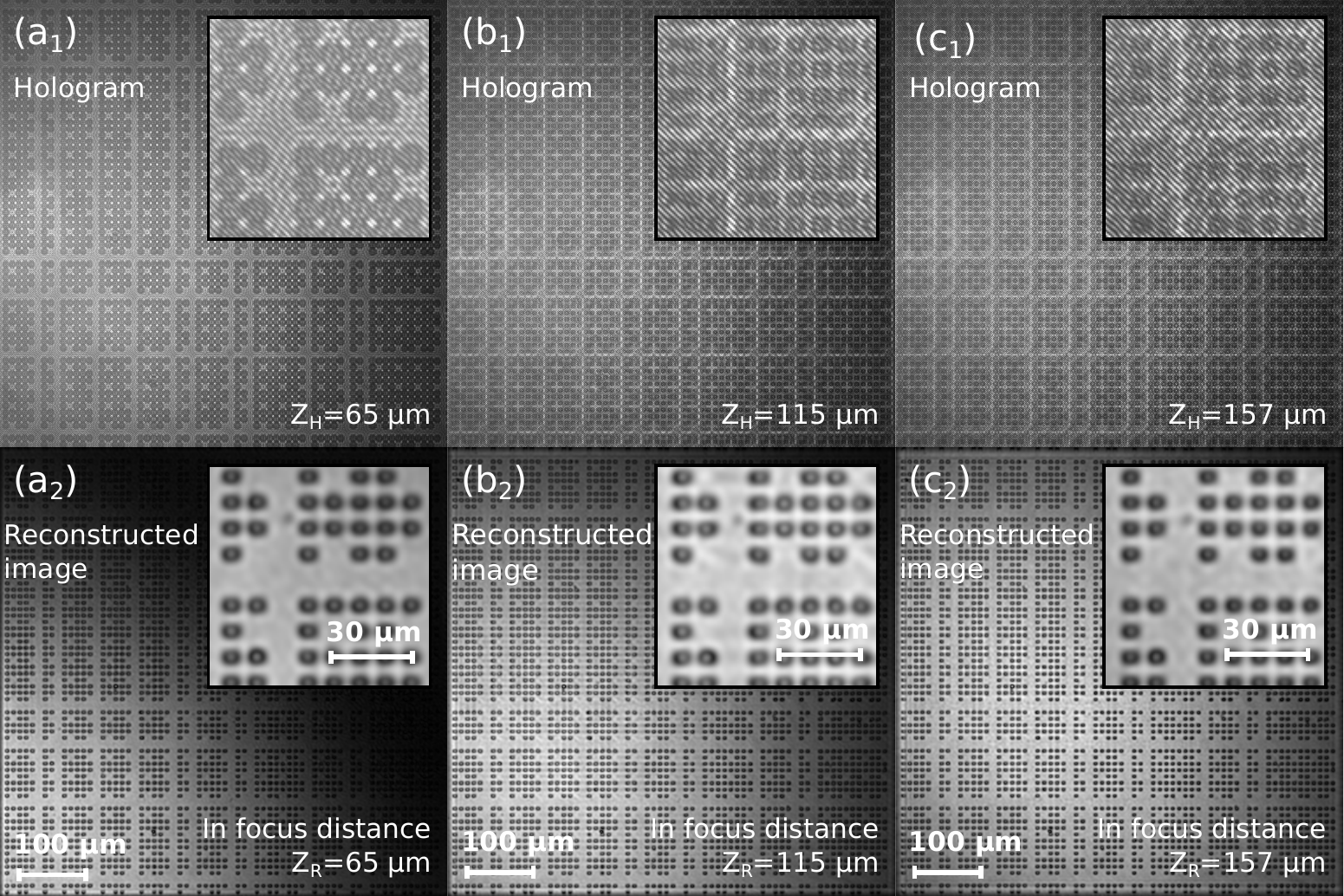}
    \caption{
    ($\mathrm{a}_1$), $(\mathrm{b}_1$) and ($\mathrm{c}_1$), experimental holograms (1024x1024 pixels) of the same area of a pseudo-periodic pattern corresponding to a propagating distance $Z_H$ of 65~µm, 115~µm and 157~µm, respectively. 
    ($\mathrm{a}_2$), ($\mathrm{b}_2$) and ($\mathrm{c}_2$), amplitude image reconstruction at a distance $Z_H = Z_R$, respectively. 
    The insets are zooms of the same sub-area of 128x128 pixels.}
    \hrulefill
    \label{fig:experimental_hologram_1}
    
\end{figure}
\section{Applications to digital holographic microscopy} \label{section:Applications to Digital Holographic Microscopy}

\subsection{Experimental setup and targeted pose measurement application} \label{subsection:Experimental setup and targeted pose measurement application}

Our final goal is to achieve 3D position and displacement measurements by means of DH combined with computer vision and thus to perform simultaneous high accuracy in-plane and out-of-plane measurements. 
For that purpose, a periodically micro-structured pattern is used in order to allow unambiguous in-plane position detection through absolute phase computations \cite{andre2020sensing}. 
Using conventional computer vision, a $10^8$ range-to-resolution ratio was demonstrated through robust phase-based decoding \cite{andre2020sensing,andre2020robust}.
However, this 2D measurement method also works with DH \cite{sandoz2011lensless,asmad2018digital}. 
In order to apply that kind of micro-structured pattern to out-of-plane motion, a DHM is used. 
This paper explores if DL, and more particularly tiny networks, are able to determine the correct focusing distance with high accuracy and robustness to speed-up the pattern intensity and phase reconstructions and to provide a more accurate Z-position estimation along an extended longitudinal direction close to 100 $\mu$m.
\\
\indent In practice, experiments were carried out on an antivibration table with a DHM (by Lyncee Tec, Switzerland) equipped with a camera with a $5.86\,\mu$m pixel size (Basler acA1920-155um), a hexapode (Newport HXP50-meca) capable of precise motions along the six degrees of freedom and a micro-encoded pattern made in our clean room facility ($2\times 2\,cm^2$, period 9~µm, 12 bits encoding \cite{andre2020sensing}) covered with a uniform 100~nm thick aluminium layer to obtain a phase object. 
This pseudo periodic pattern was observed with a MO (Leica, mag 10$\mathrm{x}$, NA$=$0.32) at wavelength $\lambda = 674.99$~nm. The light source consists of a superluminescent diode equipped with an interference filter whose width is of 5\,nm at half maximum, leading to a coherence of about 100\,{\textmu}m.
The sample was shifted along the $Z$ direction by steps of $\sim$1~µm and on a total height of $\sim92$~{\textmu}m.
At each Z step, a series of $400$ holograms ($1024\times 1024$ px, 8 bits) was recorded with random displacements along the lateral $X$ and $Y$ directions and random planar angles between $\pm8$~degrees. 
In total the experimental dataset contains 40,040 holograms.

\subsection{Autofocusing in digital holographic microscopy} \label{subsection:Autofocusing in Digital Holographic Microscopy}

The advantage of DH is to provide at different reconstruction distances $Z_R$ the complex field diffracted over a distance $Z_H$ from the hologram plane. 
The hologram propagation $Z_H$ can be tuned over an extended range of up to 92~µm in our DHM setup (limited by the light source coherence length). 
Figures~\ref{fig:experimental_hologram_1}($a_1$), ($b_1$) and ($c_1$) show an experimental hologram of 1024x1024 pixels propagated at three different distances, 65~µm, 115~µm and 157~µm, respectively.
The insets are a zoom of the same sub-area of 128x128 pixels.
Over this large propagation range, the holograms recorded are entirely different. 
Among the different reconstructed planes, the reconstruction distance $Z_R = Z_H$ corresponds to the image in focus.
Panels in Fig.~($a_2$), ($b_2$) and ($c_2$) are respectively the amplitude image reconstruction of the holograms panels in Fig.~($a_1$), ($b_1$) and ($c_1$) with a back propagation distance $Z_R = Z_H$.
The reconstruction is based on a plane waves angular spectrum method \cite{goodman1996introduction}.
Except for illumination variation along the recording distance range of 92~µm, the three reconstructed images are highly similar. 
\begin{figure}[t]
    \centering
    \includegraphics[width=0.5866\linewidth]{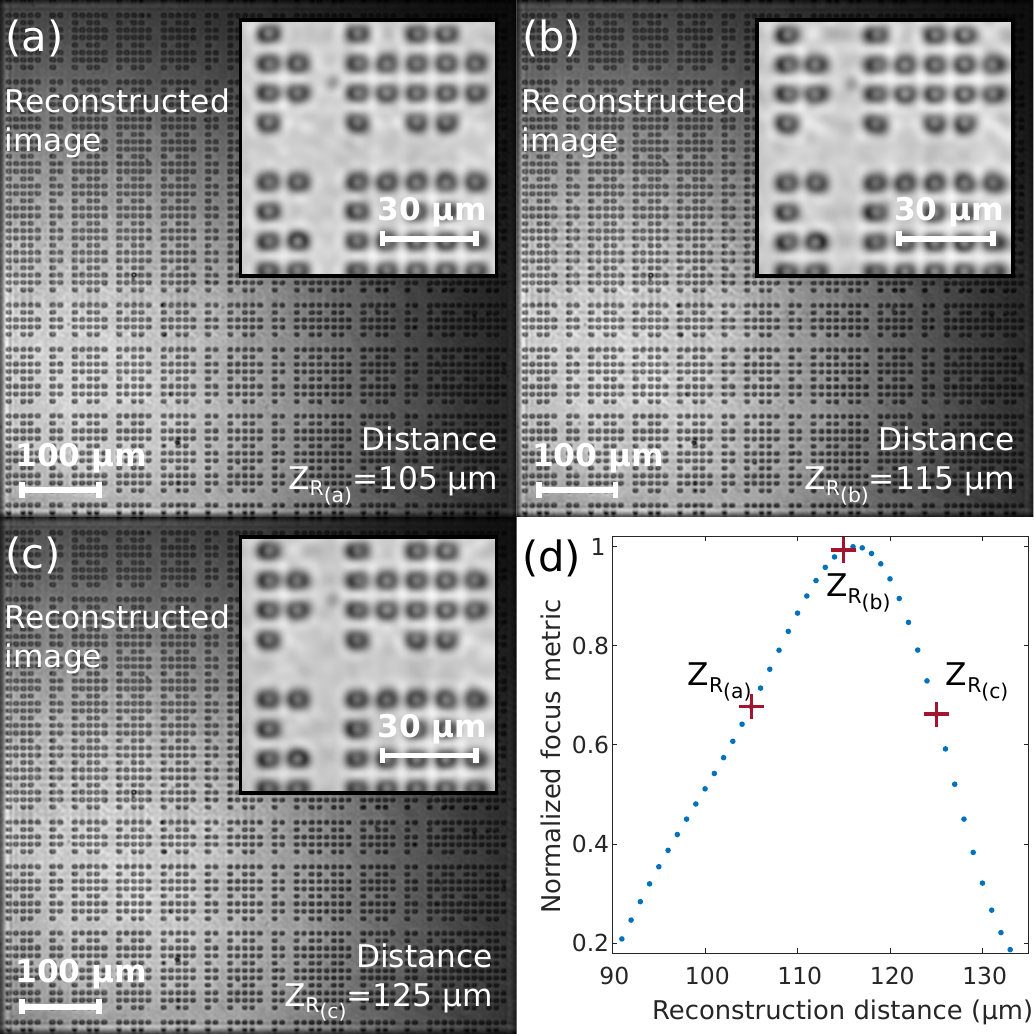}
    \caption{
    (a), (b) and (c), amplitude image reconstruction of the hologram of Fig.~\ref{fig:experimental_hologram_1}\,($\mathrm{b}_1$) at distances $Z_R$ of 105~µm, 115~µm and 125~µm, respectively.
    (d), focus estimation function calculated from the intensity image Laplacian (LAP) as described in~\cite{fonseca2016comparative}.
    The red crosses correspond to the distance reconstruction $Z_{R_{(a)}}$, $Z_{R_{(b)}}$ and $Z_{R_{(c)}}$ of the panel (a), (b) and (c), respectively.}
    \hrulefill
    \label{fig:experimental_hologram_2}
    
\end{figure}
\\
\indent To find the axial position of an object, the challenge is therefore to find this distance $Z_R$.
Autofocusing techniques in DH are applied considering the modulus of the reconstructed complex field or the modulus of spatial spectrum of the propagated field \cite{fonseca2016comparative}.
The sharpness of the image can be determined from multiple focusing criteria such as the sum of the modulus of the complex field amplitude, the use of a logarithmically weighted Fourier spectral function, the variance of gray value distribution, focus measure based on autocorrelation, absolute gradient operator, Laplace filtering, Tamura coefficient estimation, or wavelet-based approaches \cite{fonseca2016comparative}.
A comparison of many focusing criteria in terms of computational time and accuracy in determining the focal plane have been already discussed \cite{fonseca2016comparative, langehanenberg2008autofocusing}. 
\\
When the focus criterion is at an extremum, the focus of the reconstructed image is optimal. 
There are also various methods that would allow an automated determination of the optimal reconstruction distance \cite{dougar2013real}. 
However, all these approaches require the numerical reconstruction of a set of images within a given range of propagation distances. 
Then, the focus criteria is calculated from each reconstructed image, to determine the distance of focusing. 
Figures~\ref{fig:experimental_hologram_2}(a), (b) and (c) show three reconstructed images at different distance $Z_R$ of the experimental hologram of the Fig.~\ref{fig:experimental_hologram_1}($b_1$).

These three images spaced by 10~µm from each other illustrate the difficulties of obtaining a sharp focus criteria.
Figure~\ref{fig:experimental_hologram_2}(d) shows the result of a focus metric based on the image Laplacian~\cite{fonseca2016comparative} where the red crosses correspond to the case of the panels (a), (b) and (c).
The resolution of this normalized autofocusing method is close to the DoF provided  by the Numerical Aperture ($\mathrm{NA}$) of the MO and the wavelength $\lambda$ used, which gives in our case $\mathrm{DoF}=\sfrac{2\lambda}{\mathrm{NA}^2}=15$~µm.
Even if these approaches could be efficient \cite{dougar2013real}, they are computationally demanding and time consuming, especially if the size of the hologram is large.

\section{Results}

This section shows the results obtained by running a series of inferences on test sets of different holograms. All the test results have been generated using the proposed tiny models: TViT, TVGG and TSwinT.
Four datasets are considered: experimental and simulated phase holograms of the pseudo-periodic pattern, and amplitude and phase holograms of a simulated USAF pattern. 
The simulated holograms are generated by using a plane-wave spectrum propagation algorithm. Although the simulation reproduces the experimental parameters of the sample and the imaging system, it is deliberately free of motion uncertainties, surface defects, optical aberration and noise.
This approach allows to obtain the intrinsic performance limit of the neural networks proposed. 
Free of mechanical limitation, the simulated Z pitches are therefore less than 1~µm and over a total range of 100~µm.

\indent For each set of holograms, TViT, TVGG and TSwinT have been trained from scratch. 
The neural networks have been configured to apply a regression on the input data. 
A total of 200 epochs have been executed and the learning curves have correctly converged. 
The learning rate was set to $1\cdot 10^{-4}$ using the Adam optimizer \cite{kingma2017adam}. 
During the training, a total of 64 (TViT) or 32 (TSwinT \& TVGG) ROIs are selected randomly for each hologram.
As showed in \cite{chen2019log}, the $\mathrm{log}(\mathrm{cosh})$ loss function can improve the result of Varational Auto-Encoder. 
This loss function 
\begin{equation}
    L(Z_H, Z^{\mathrm{Pred}}) = \sum_{i=1}^n \mathrm{log}(\mathrm{cosh}(Z^{\mathrm{Pred}}_i - Z_{H_i})),
    \label{eq:log_cosh_loss_function}
\end{equation}
is also less prone to outliers than the mean squared error (MSE) or the mean absolute error (MAE) where $n$ is the number of training samples, $Z$ and $Z^{\mathrm{Pred}}$ are the expected and predicted values, respectively.
\\
Table~\ref{tab:validation_loss_comaparison} shows the performance of the validation loss functions on our experimental hologram dataset for all tiny networks. 
Table \ref{tab:validation_loss} shows the best validation loss $L$ for each model and each training set.
In the following sections, the error $\epsilon$ of one inference is measured by,
\begin{equation}
    \epsilon = Z_R^{\mathrm{Pred}}-Z_H.
\end{equation}
All the codes to train the proposed tiny networks (TViT, TVGG and TSwinT)  are accessible at this address \url{https://github.com/scuenat/DHMTinyNetworks} and the data is available in \cite{ExperimentalDataset}.

\begin{table}[h!]
    \centering
    \begin{adjustbox}{max width=\columnwidth}
    \begin{tabular}{c|c|c|c|}
        \textbf{Model}  & \textbf{MSE} & \textbf{MAE} & \textbf{log cosh}\\
        \hline
        TVGG & 0.25 & 0.39  & 0.11 \\
        TViT & 0.42 & 0.50 & 0.13 \\ 
        TSwinT & 0.30  & 0.43 & 0.12  \\
    \end{tabular}
    \end{adjustbox}
    \caption{Comparison of the validation loss functions (log cosh, MSE or MAE) for each proposed tiny models trained on experimental holograms.}
    \hrulefill
    \label{tab:validation_loss_comaparison}
\end{table}

\begin{table}[h!]
    \centering
    \begin{adjustbox}{max width=\columnwidth}
    \begin{tabular}{c|c|c|c|c|}
        \textbf{Model} &  \multicolumn{2}{c|}{\textbf{Pseudo-periodic pattern}} & \multicolumn{2}{c|}{\textbf{USAF (simulated)}}  \\
         \cline{2-5}
    & \textbf{Experimental} & \textbf{Simulated} & \textbf{Amplitude} & \textbf{Phase}\\
    \hline
        TVGG & 0.11 & 0.003 & 0.04 & 0.05  \\
        TViT & 0.13 & 0.004 & 0.09 & 0.04 \\
        TSwinT & 0.12  & 0.012  & 0.05 & 0.07 \\
    \end{tabular}
    \end{adjustbox}
    \caption{Validation loss (log cosh) for each model and each set of holograms.}
    \hrulefill
    \label{tab:validation_loss}
\end{table}

\begin{figure}[!t]
    \centering
    \includegraphics[width=250pt]{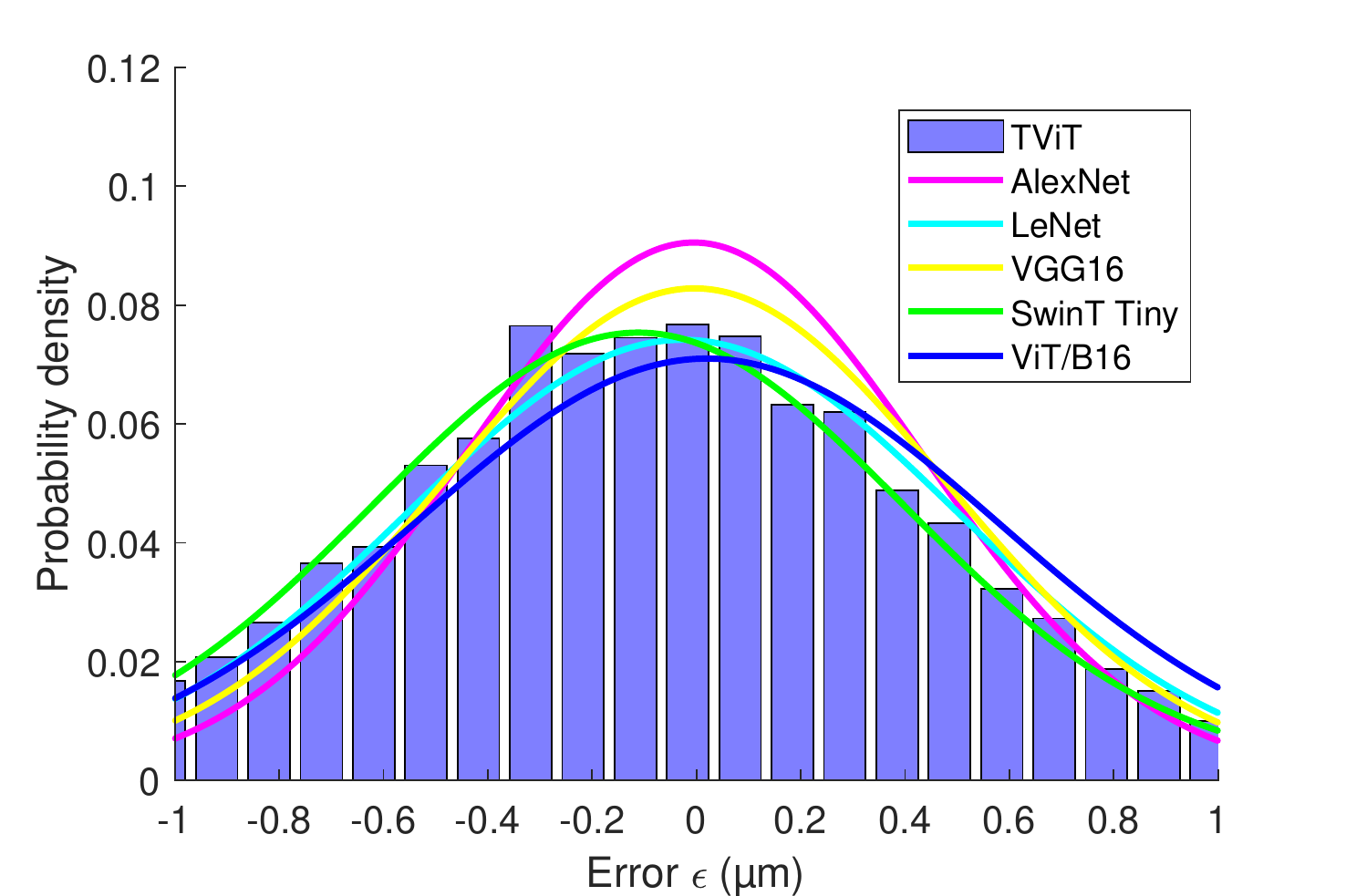}
    \caption{Error distribution comparison between TViT and other state-of-the-art models like AlexNet, LeNet and original models such as VGG16, SwinT Tiny and ViT/B16. 
    $\bar{\varepsilon}_{\mathrm{TViT}} = -0.07 \pm 0.61~\mathrm{{\mu}m}$,
    $\bar{\varepsilon}_{\mathrm{AlexNet}} = 0 \pm 0.51~\mathrm{{\mu}m}$, $\bar{\varepsilon}_{\mathrm{LeNet}} = -0.02 \pm 0.61~\mathrm{{\mu}m}$,
    $\bar{\varepsilon}_{\mathrm{VGG16}} = 0 \pm 0.56~\mathrm{{\mu}m}$,
    $\bar{\varepsilon}_{\mathrm{SwinT~Tiny}} = -0.17 \pm 0.59~\mathrm{{\mu}m}$,
    $\bar{\varepsilon}_{\mathrm{ViT/B16}} = 0.01 \pm 0.66~\mathrm{{\mu}m}$},
    \hrulefill
    \label{fig:all_models_comparison}
\end{figure}

\subsection{Experimental holograms: pseudo periodic pattern}

\begin{figure}[!b]
    \centering
    \includegraphics[width=300pt]{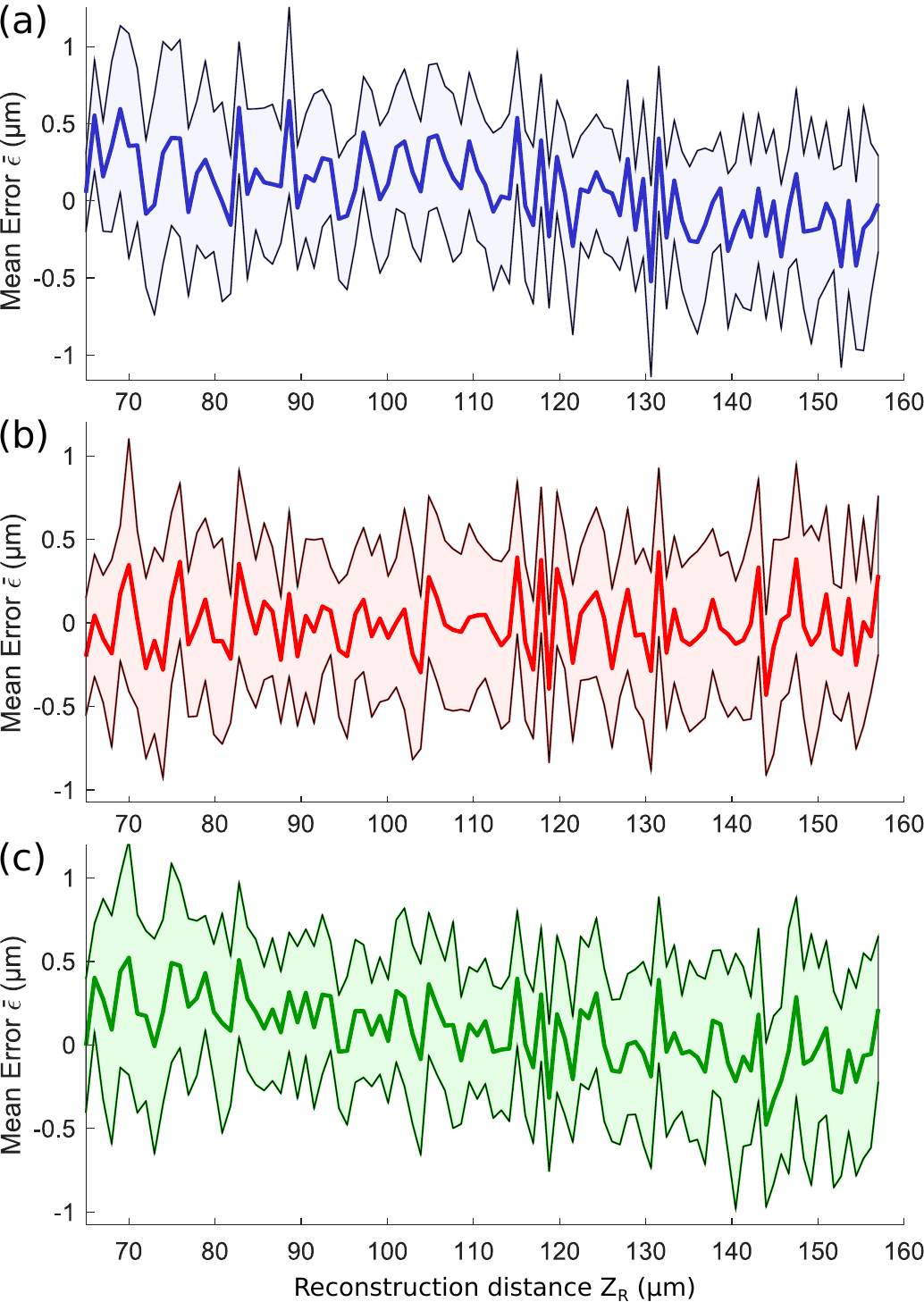}
    \caption{Prediction error on experimental holograms of pseudo-periodic patterns. 
    The bold lines (area) are the average (standard deviation) of the error for each reconstruction distance over 92~{\textmu}m. 
    (a), (b) and (c) correspond to TViT, TVGG and TSwinT models, respectively.}
    \hrulefill
    \label{fig:experimental_inference_error}
\end{figure}

Holograms of Fig.~\ref{fig:experimental_hologram_1}\,(a$_1$),\,(b$_1$)\,and\,(c$_1$), recorded at different distances $Z_H$, are representative specimens of the set of experimental holograms used by the tiny networks during the training phase.
The experimental dataset, which contains a total of 40,400 holograms (400 holograms for each distance $Z_H$), was distributed between learning, validation and testing sets with a $70/20/10$ ratio. 
The models have therefore been tested on a set of 4,040 holograms, 40 holograms for each $Z_H$ spaced by 1.0~{\textmu}m ranging on 92~{\textmu}m.
In Fig.\,\ref{fig:experimental_inference_error}, the results of the inferences testing of the TViT, TVGG and TSwinT are represented on the panels (a), (b) and (c), respectively.
The average and the Full Width at Half Maximum (FWHM) of the error are represented by the color bold curves and areas, respectively.
Comparable performances for the three neural networks with a high stability along the full range and small errors can be observed.  
Figure~\ref{fig:error_distribution} gives another view allowing to appreciate the error distribution.
The solid lines are Gaussian fits of the error for each network model.
The average and standard deviation (half of the FWHM) are given for each case.
Panel~(a) illustrates the results of the experimental dataset. 
This graph shows an error bounded by 1~{\textmu}m for all models. Figure \ref{fig:all_models_comparison} compares the error distribution of a TViT model with reference neural networks in digital holography as VGG16, LeNet (as presented in \cite{Jaferzadeh:19}) and AlexNet (as presented in \cite{pitkaaho2019focus}). The original versions, SwinTransformer Tiny (SwinT Tiny) and ViT/B16, are also represented.
For the MO used, the 1 µm autofocusing accuracy achieved is 15 times smaller than the theoretical DoF.

\subsection{Simulated holograms: pseudo periodic pattern}
\begin{figure}[t!]
    \centering
    {\includegraphics[width=1\textwidth]{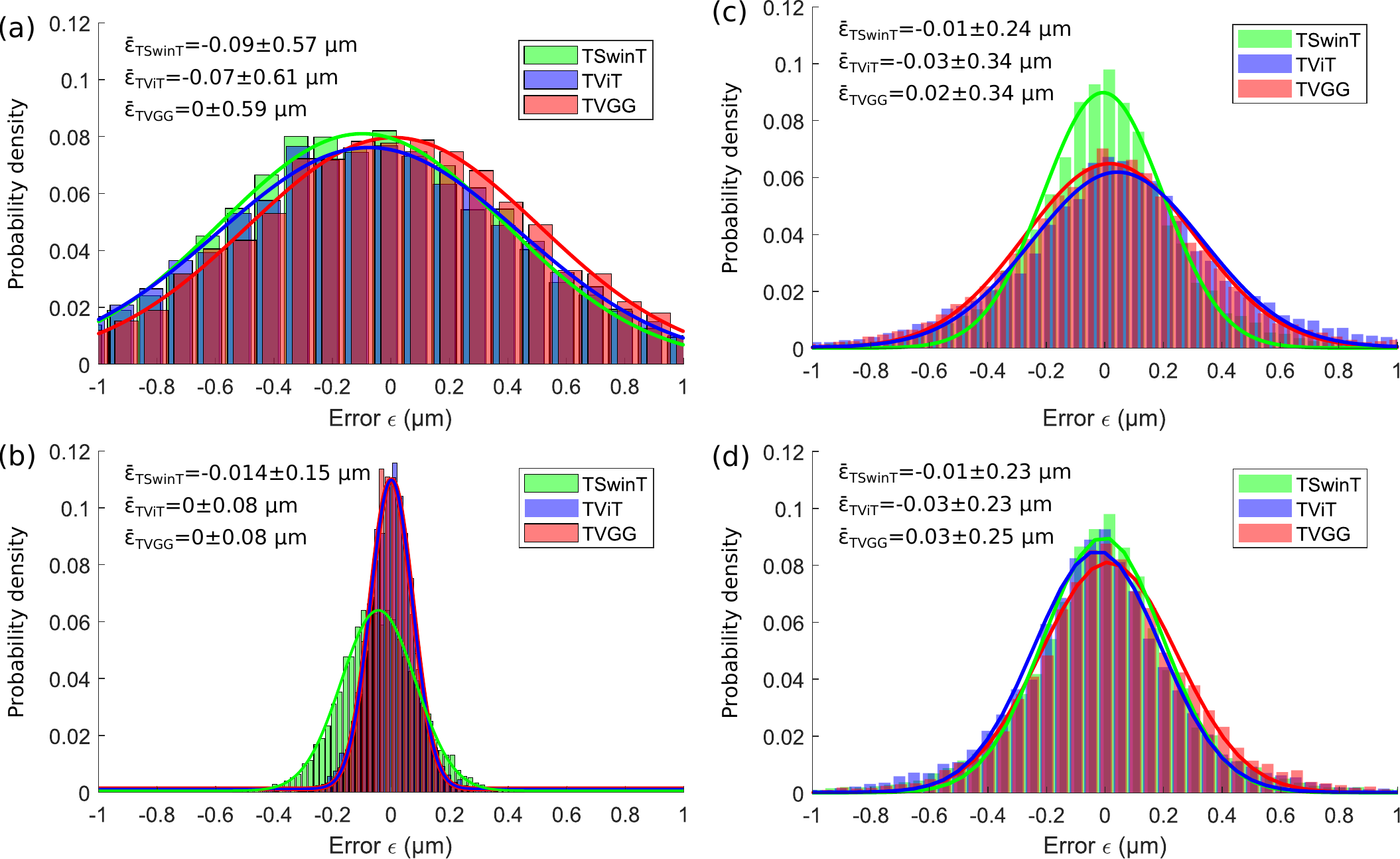}}
    \caption{
    Error distribution of the three neural networks over the reconstruction distance Z. 
    (a) and (b) are the results for the experimental and simulated pseudo-periodic patterns, respectively.
    (c) and (d) are the results for the simulated USAF patterns in amplitude and phase, respectively.
    }
    \hrulefill
    \label{fig:error_distribution}
\end{figure}

Simulated phase holograms of pseudo-periodic patterns are used to test the limit of the proposed models performances.
40,040 holograms, including 40 different sites (in-plane position and orientation) vertically scanned per steps of 1\,$\mu$m, ranging on 100~µm, constitute the full training dataset.
The testing dataset is composed of 6,819 holograms never viewed, spaced by 0.1~µm and ranging on 100~µm. 
Figure~\ref{fig:error_distribution}(b) shows the error distribution of the reconstruction distance $Z_R$ for the three models. 
Without noise and exact Z position labelling, the reconstruction error considerably decreases below 0.3~µm.  
The best models are TViT and TVGG which have a FWHM of the error distribution of 0.160~µm.
These great performances for a testing with a step size ten time smaller than the training dataset prove the high regression quality of the three tiny networks.

\begin{figure}[t!]
    \centering
    {\includegraphics[width=250pt]{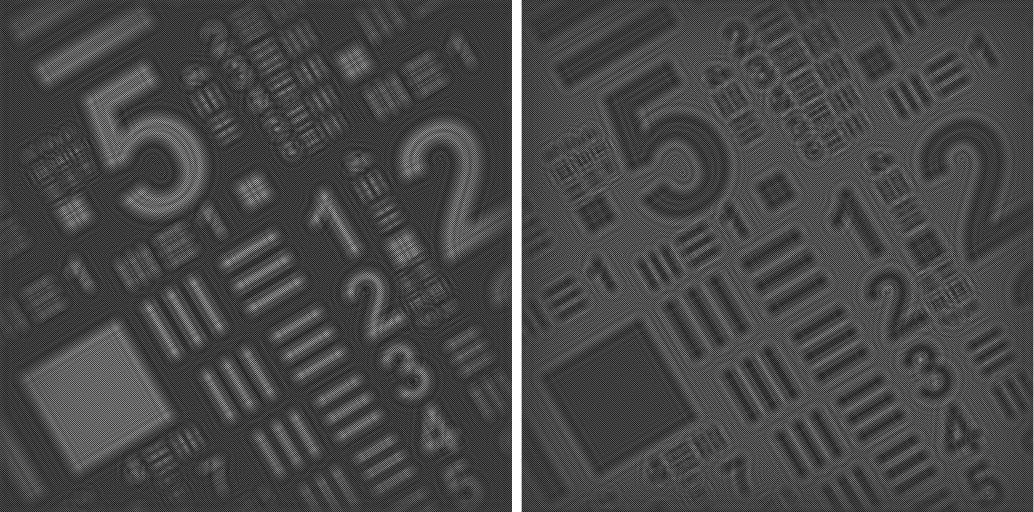}}
    \caption{
    (a) and (b), two representative examples of simulated 1024X1024 USAF hologram in amplitude and phase, respectively.
    } 
    \hrulefill
    \label{fig:usaf_holograms}
\end{figure}

\begin{figure}[b!]
    \centering
    \includegraphics[width=400pt]{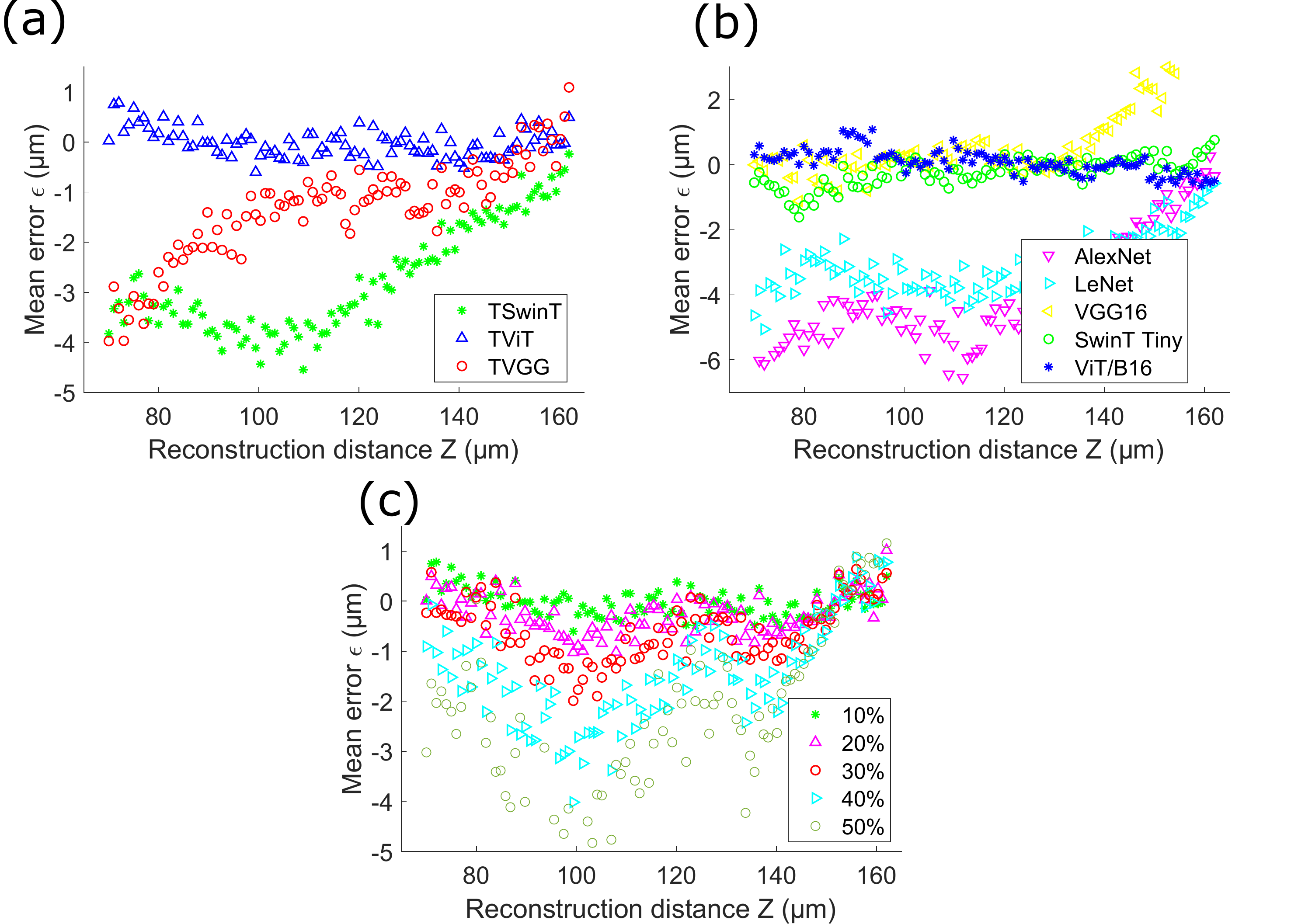} 
    \caption{
    Experimental results for all models in the case where the network input ROI artificially undergoes a loss of information.
    (a) shows the average error over the entire Z reconstruction range for the proposed tiny models with a loss of 10\%, (b) shows the same but for original and reference models and (c) the limit of occlusion for a TViT model}.
    \hrulefill
    \label{fig:occlusion_resistance}
\end{figure}
\subsection{Amplitude and phase object: USAF 1951 resolution chart}
In comparison with the pseudo-periodic sample, the USAF pattern is a more complex pattern due to a wider spatial frequencies bandwidth. 
To avoid building a dataset with empty regions without information, the pattern was simulated by filling the space more densely than a commercial target.
In order to characterize the tiny networks performances in function of the amplitude or phase nature of the holograms, two different datasets were constituted.
Figures~\ref{fig:usaf_holograms}(a) and (b) show two representative holograms in amplitude and phase, respectively. 
Both training (testing) datasets are constitued of 400 (50) different holograms at each step of 0.5\,$\mu$m ranging on 130~µm (100~µm), for a total of 104,400 (10050) holograms.
Figures~\ref{fig:error_distribution}(c) and (d) show the error distribution of the tiny networks in case of amplitude and phase objects, respectively.
In comparison with the pseudo-periodic pattern, all tiny networks have worse performances but stays highly competitive with a error below 0.35~µm.

In case of amplitude holograms, the TSwinT model shows better results than the other two.
Contrary to the TSwinT model which has the same performances whatever the characteristic of the hologram, the TViT and the TVGG give more precise inferences for the phase holograms.
In this case, the three neural networks show similar performances with an error smaller than 0.25~µm.

\subsection{Occlusion test}
Another type of neural network performance is its robustness in a degraded configuration.
Experimentally we have already seen that the three proposed models are resilient with respect to homogeneous sample illumination, as shown in Fig.~\ref{fig:experimental_hologram_1} and Fig.~\ref{fig:experimental_hologram_2}.
To go further, areas of the experimental testing dataset have been deleted.
A random squared region of 10\% of the ROI selected as input for the tiny neural networks is uniformly set to zero. 
This operation simulates a dust or a sample defect and evaluates the degree of locality of the neural networks.
The results obtained are shown in Fig.~\ref{fig:occlusion_resistance}, where panel~(a) display the error average along the Z for the proposed tiny models, panel~(b) display the error average along the Z for the reference and original models and panel~(c) the limit of occlusion in case of a TViT model.
It can be observed that the TSwinT and TVGG models are highly impacted by 10\% of occlusion as the models, AlexNet and LeNet. In contrast, TViT is clearly the most robust architecture against occlusion. Figure~\ref{fig:occlusion_resistance}(a) shows that on average the TViT error remains stable on the full 92\,µm range.

\subsection{Inference speed}

\begin{figure}[!h]
    \centering
    \includegraphics[width=300pt]{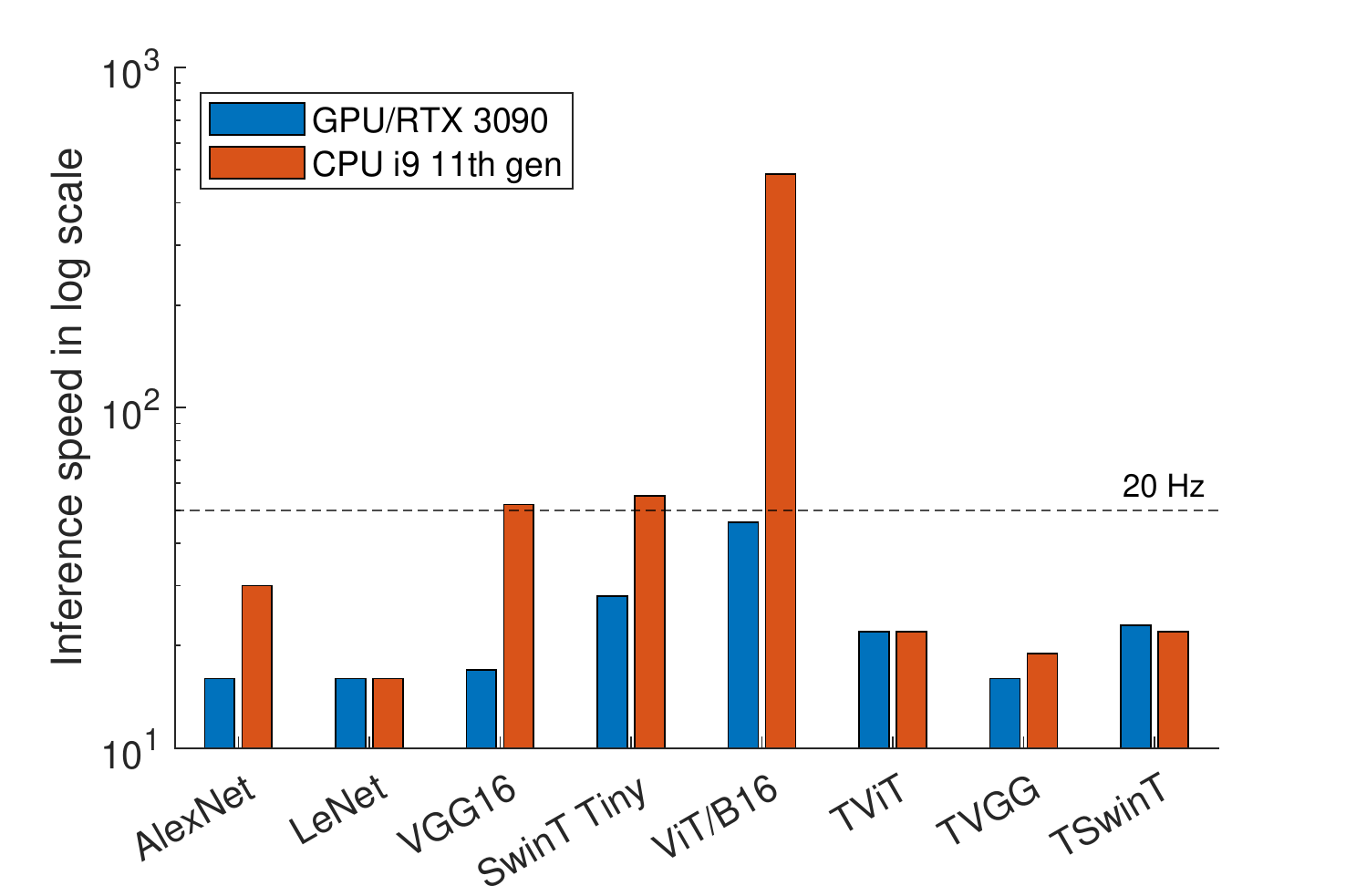}
    \caption{
    Comparison of the inference speed in log scale for AlexNet, LeNet, VGG16, SwinT Tiny, ViT/B16, TVGG, TSwinT and TViT on different architectures; GPU: RTX 3090 24Gb and CPU: Intel i9-11900K @3.50GHz.The dashed line represents the real-time limit in robotics.}
    \label{fig:inference_speed_tVit}
    \hrulefill
\end{figure}

Whether for applications in microrobotics or in 3D microscopy for life sciences, there is great interest in being able to work in real time and with commercially accessible equipment.
Therefore, Figure~\ref{fig:inference_speed_tVit} shows the comparison of the median speed of 200 inferences with two different configurations for all the neural networks compared (AlexNet, LeNet, VGG16, SwinT Tiny, ViT/B16, TViT, TVGG and TSwinT).
On the Intel i9-11900K @3.50GHz CPU the performance is comparable to using a GPU NVidia RTX 3090, 24Gb, with an inference speed below 25~ms for LeNet and TViT.
As analyze in details \cite{pitkaaho2019focus}, the reconstruction time of an hologram for twenty different distances takes a total of 318 ms on an Intel Core i5 processor.
The image reconstruction knowing the predicted distance $Z_R^{\mathrm{Pred}}$ is therefore of $\sim15$~ms.
This value is also confirmed by the DHM which has a reconstruction rate of up to 60 frames per second.
The tiny models proposed with low inference times, associated with an image reconstruction algorithm, therefore form a solution compatible with the constraints of real applications. 

\section{Discussion \& conclusion}
TVGG, TViT and TSwinT give close results when taking the different sets of holograms (pseudo-periodic pattern experimental/simulated and simulated USAF phase/amplitude). 
In this paper, it has been shown that TViT is more robust in presence of occlusion (Fig.~\ref{fig:occlusion_resistance}), considering that Z depth information is present on the entire hologram due to diffraction properties in coherent imaging.
A TViT model benefits from the multi-head self-attention (Fig.~\ref{fig:multi_head_attention}) which  takes the complete ROI at each layer in consideration (not a set of extracted features). 
A CNN like TVGG works a bit differently as it tries to build a set of features through its first Convolution/Pooling layers followed by full connected layers (regression). 
A CNN, by extracting at each layer a more complex representation of the features, explains why it  focuses on dedicated regions which impacts the accuracy of the inference in presence of occlusion.
According to the above, a TViT model seems more suited to the prediction of the in-focus distance in DHM, as it scans everywhere and is more robust in terms of occlusion. 
This goes in the same direction as presented in \cite{naseer2021intriguing,cuenat2021convolutional} where it has been shown that a ViT model is a lot more robust than a CNN. 
Although, a TSwinT model is based on the derivative of a ViT model, it does not perform as well as a TViT in case of added occlusion. 
As TSwinT is only applying the self-attention on a set of windows (W-MSA) or shifted windows (SW-MSA) (through its Swin-Transformer Blocks, Fig. \ref{fig:tiny_vit_architecture}(c2)), it can be assumed that the occlusion has a bigger impact on the result of the multi-head attention than a pure ViT like TViT. 

Figure~\ref{fig:all_models_comparison} shows that the tiny models (TVGG, TSwinT and TViT) are as accurate as the original versions (VGG16, SwinT Tiny and ViT/B16). It also shows that a AlexNet or LeNet model reach similar performance. Considering the inference speed on CPU (Figure~\ref{fig:inference_speed_tVit}) and the robustness against an occlusion, TViT is the best model proposed.

As mentioned in \cite{raghu2021vision}, a ViT (TViT) would need a lot of data to be trained from scratch. 
This is not what has been experienced, as a TViT can be trained from scratch using our set of experimental holograms of pseudo-periodic pattern using a total of 2,327,040 ROIs (36,360 holograms $\times$ 64 ROI). 
A huge amount of data is normally needed as a ViT (TViT) projects each patch on an embedded dimension (Fig.~\ref{fig:tiny_vit_architecture}(a), Patch embedding). 
The reconstruction distance Z information is spread over the complete space of the hologram, which is most likely an argument to explain why a ViT-like network can learn from scratch without having a huge dataset at disposal.

Our experiments showed that the reconstruction distance Z can be predicted in DHM with a high accuracy using deep learning last generation techniques, especially regression models. 
An error bounded by $\sim$1~{\textmu}m on the reconstruction distance Z has been reached for a dataset of experimental holograms on a range of 92~{\textmu}m. 
The regression approach allows experimentally to surpass the DoF of the MO by an order of magnitude.
Moreover, this error can be lowered down to $\sim$0.3~{\textmu}m when the models are trained on the simulated holograms of a pseudo-periodic pattern or USAF pattern (phase or amplitude).
The discrepancy between results obtained from experimental and simulated datasets is partly due to the limited accuracy of the actuator used where bi-directional repeatability is of 0.3~µm. Acquisition noise may also play a significant role in that reduction of performances obtained from experimental datasets.
All proposed tiny models offer an alternative to expensive GPUs as the time for an inference is below the real-time limit in robotics of 20 Hz (Fig.~\ref{fig:inference_speed_tVit}), less than 25~ms on an Intel i9.

The ability of tiny networks to determine the in-focus depth with a FWHM of about one micron opens attractive application prospects. 
Indeed, if two wavelength DHM are considered, the ambiguity range is about twice the FWHM demonstrated in this paper (with our commercial DHM, $\lambda_1 = 674.99$~nm and $\lambda_2=793.63$~nm; i.e. $\lambda_{eq}/2=2.25$~µm). 
This means that DL may bridge the gap between the MO DoF and the Z-information provided by the interferometric phase to achieve Z-position determination down to the interference sensitivity; i.e. around 1~nm over ranges of tens of microns. 
Such a prospect would significantly improve the current capabilities of computer vision position sensing applied to 3D microscopy.

\section{Acknowledgments}
This work was supported by HOLO-CONTROL (ANR-21-CE42-0009), TIRREX (ANR-21-ESRE-0015), SMARTLIGHT (ANR-21-ESRE-0040) by Cross-disciplinary Research (EIPHI) Graduate School (contract ANR-17-EURE-0002), Région Bourgogne Franche-Comté (HoloNET project).
This work was performed using HPC resources from GENCI-IDRIS (Grant 20XX-AD011012913) and also the Mésocentre de Franche-Comté.

\section*{Disclosures}
The authors declare no conflicts of interest.

\section*{Data availability} Data underlying the results presented in this paper are available in Zenodo, Ref.~\cite{ExperimentalDataset}.

\bibliographystyle{unsrt}  
\bibliography{references}

\end{document}